\documentclass{article}

\usepackage[utf8]{inputenc}
\usepackage[T1]{fontenc}
\usepackage{lmodern}
\usepackage[margin=1in]{geometry}
\usepackage{float}
\usepackage{mathtools}
\usepackage{amsmath}
\usepackage{amssymb}
\usepackage{amsfonts}
\usepackage{bbm}
\usepackage{bm}
\usepackage{graphicx}
\usepackage{subcaption}
\usepackage{caption}
\usepackage{wrapfig}
\usepackage{tikz}
\usepackage{pgfgantt}
\usepackage{booktabs}
\usepackage{soul}
\usepackage{url}
\usepackage{pdfpages}
\usepackage[title]{appendix}
\usepackage{siunitx}
\usepackage{algorithmic,algorithm}
\usepackage{setspace}
\usepackage{hyperref}
\DeclareSIUnit\foot{ft}
\DeclareSIUnit\sqft{\foot\squared}

\title{Cost and Complexity as Barriers to RTLS Adoption in SMEs:\\
A Survey and Analysis}
\author{
    Peyman Moeini\thanks{Email: peyman.moeini@mail.mcgill.ca},
    Mark Coates\thanks{Email: mark.coates@mcgill.ca} \\
    Department of Electrical and Computer Engineering \\
    McGill University, Montreal, Canada
}

\begin{document}
\maketitle

\tableofcontents
\newpage

\begin{abstract}
Real-time location systems (RTLSs) are a core enabling technology for Industry~4.0 and emerging Industry~5.0, providing the spatiotemporal data required for asset tracking, workflow optimization, safety, and integration with Warehouse Management Systems (WMS), Manufacturing Execution Systems (MES), and digital twins. Although RTLSs are increasingly deployed by large enterprises, adoption among small and medium-sized enterprises (SMEs) remains limited. This paper investigates whether \emph{cost} and \emph{installation complexity} constitute primary barriers to RTLS adoption in SME contexts.

We first situate RTLSs within the broader Industry~4.0 and Logistics~4.0 landscape and summarize their value proposition for industrial operations. We then synthesize evidence from the literature on technical, financial, and organizational constraints, with particular emphasis on infrastructure requirements, calibration effort, integration with legacy systems, and human factors. To complement this qualitative analysis, we report results from an online survey of sixteen manufacturing and technology professionals in Canada and the United States. Respondents report high perceived operational value for real-time tracking, but identify upfront cost, installation effort, integration with existing systems, and dependence on multiple anchor nodes as dominant obstacles. A majority indicate acceptable upfront investment levels below \$10{,}000 and express a strong preference for low-infrastructure deployments with minimized anchor counts.

Building on these findings, we discuss design directions for RTLSs targeted at SMEs, including wireless and modular architectures, cloud-managed and self-calibrating solutions, standardized integration interfaces, and anchor-minimizing or anchor-free localization approaches. The analysis highlights that RTLS adoption in SMEs is constrained less by a lack of perceived value than by misalignment between current deployment models and SME resource and infrastructure constraints.
\end{abstract}

\section{Introduction}
\subsection{Background and Motivation}
The Fourth Industrial Revolution, commonly referred to as Industry~4.0, denotes the progressive digitalization of manufacturing and industrial operations through the integration of cyber-physical systems (CPS), the Industrial Internet of Things (IIoT), and real-time data analytics \cite{szaszIndustry40Review2020, siagianSupplyChainIntegration2021}. Sensors, controllers, and information systems are increasingly interconnected, enabling predictive maintenance, automated material handling, and data-driven decision-making at the shop-floor level \cite{jangAreSmartManufacturing2022}. Within this evolving landscape, RTLSs play a central role by generating continuous, spatially resolved information on the movement and status of assets, materials, and personnel \cite{daher2024rtls}.

While RTLSs are increasingly recognized as an essential component of digital factories and smart warehouses, most detailed case studies and commercial deployments focus on large enterprises with considerable capital resources and dedicated IT and operations-technology (OT) teams. In contrast, SMEs, which constitute the majority of firms in many industrial economies and play a central role in innovation and employment \cite{kumar2025msme}, often operate with strict budgetary constraints, legacy infrastructure, and limited internal technical capacity \cite{shahimi2025aiadoption}. It is therefore not obvious that RTLS solutions and deployment models designed for large facilities transfer directly to SME contexts.

\subsection{Research Question and Contributions}
This paper addresses the following overarching research question:
\begin{quote}
\emph{To what extent do the cost and installation complexity of contemporary RTLS solutions act as structural barriers to adoption in small and medium-sized industrial enterprises?}
\end{quote}

To answer this question, we make three contributions:
\begin{enumerate}
    \item We provide a structured overview of RTLS architectures and their role in Industry~4.0 and Logistics~4.0, focusing on how RTLS-generated spatiotemporal data is integrated into higher-level systems such as WMS, MES, and digital twins.
    \item We synthesize evidence from the literature on the technical, financial, and organizational challenges associated with RTLS deployment, with particular emphasis on factors that are likely to be more severe in SME settings: infrastructure constraints, calibration and maintenance requirements, integration with legacy systems, and human factors.
    \item We complement the literature review with an online survey of sixteen manufacturing and technology professionals in Canada and the United States. The survey quantifies perceptions of RTLS value, acceptable cost levels, and attitudes toward infrastructure intensity and anchor-node deployment, with a view toward identifying design directions for SME-oriented RTLS solutions.
\end{enumerate}

The combined analysis shows that practitioners broadly recognize the operational value of RTLS, yet perceive implementation cost and complexity, driven largely by installation, integration, and calibration effort, as a major barrier. The survey results further indicate that many potential adopters operate under relatively tight capital-expenditure constraints and strongly prefer low-infrastructure architectures with minimized anchor counts. These findings motivate a discussion of emerging design approaches, including anchor-minimizing and anchor-free localization frameworks, that may be better aligned with SME requirements.

\subsection{Structure of the Paper}
The remainder of the paper is organized as follows. Section~2 introduces RTLS architectures and their role in Industry~4.0 and Logistics~4.0, and outlines key structural deployment challenges. Section~3 examines the operational value and limitations of RTLS in industrial environments, distinguishing between raw location data and actionable decision support. Section~4 focuses on adoption barriers specific to SMEs, highlighting economic, infrastructural, and organizational constraints. Section~5 presents and analyzes survey results on practitioner perceptions of RTLS cost, complexity, and infrastructure requirements. Section~6 discusses opportunities for innovation in SME-oriented RTLS design and deployment. Section~7 concludes and outlines directions for future research.

\section{RTLS in Industry 4.0 and Logistics 4.0}
\subsection{RTLS Architectures and Integration with Industrial Systems}
An RTLS determines and reports the position of tagged assets or personnel within a defined physical space. Most RTLS implementations share three core components: (i) \emph{tags} attached to objects or individuals, (ii) fixed \emph{anchors} or \emph{readers} that receive signals from those tags, and (iii) a \emph{localization engine} that converts raw measurements such as time-of-flight, received signal strength, or angle-of-arrival into real-time coordinates \cite{li2009multifrequencyRFID, farid2013indoorLocalization}. The resulting time-stamped trajectories are consumed by higher-level platforms (e.g., WMS, MES, or digital twins) to generate operational insights, trigger alerts, and orchestrate automated workflows \cite{AT2025tracking}.

RTLSs rely on a variety of physical-layer technologies, including ultra-wideband (UWB), Bluetooth Low Energy (BLE), Wi-Fi, RFID, infrared, and hybrid schemes \cite{salzmann2025framework}. Each modality exhibits distinct trade-offs between accuracy, range, cost, energy consumption, and robustness to multipath and non-line-of-sight conditions \cite{sidiropoulos2025implementing, lachvajderova2024real}. UWB systems, for example, can provide centimetre-level accuracy and strong non-line-of-sight performance, but typically require precise anchor placement and synchronization \cite{vanherbruggen2024twr, barbieri2020uwb}. BLE-based solutions tend to be less accurate but may be more attractive where low cost and battery longevity are primary design drivers.

In modern deployments, RTLSs are tightly integrated into IIoT and CPS frameworks, streaming high-resolution spatiotemporal data for analytics, predictive modelling, and process automation \cite{mp2025lora}. RTLS data is ingested by MES and WMS platforms for workflow orchestration and by digital twins that represent production lines and warehouses \cite{sidiropoulos2025implementing, li2024indoor, behnke2023real, aouani2023digitalTwinWarehouse}. As a consequence, RTLSs increasingly function as a foundational sensing layer within multi-tier industrial information architectures rather than as isolated systems.

\subsection{RTLS in Logistics 4.0 and Emerging Industry 5.0}
In supply-chain and logistics operations, RTLSs contribute to the broader concept of \emph{Logistics~4.0} by enabling end-to-end traceability of goods and vehicles across networks of warehouses, production facilities, and transportation hubs \cite{tijan2019blockchainLogistics}. Real-time geolocation data supports synchronized material flows, improves delivery accuracy, and strengthens resilience to disruptions. RTLS-enabled logistics networks can dynamically optimize routing and scheduling under fluctuating demand and capacity conditions, thereby reducing operational delays and energy consumption \cite{zhang2024decision}. When integrated with digital twins, RTLS data allows continuous monitoring of asset status, location, and environment, supporting transparent and auditable supply chains \cite{behnke2023real}.

Industry~5.0 extends the Industry~4.0 vision by emphasizing human-centric, sustainable, and resilient production systems \cite{ivanov2022industry50framework}. In this paradigm, RTLSs provide contextual information that supports human--machine collaboration, adaptive work environments, and advanced safety management \cite{slovak2021visionRTLSsafety}. Human-centric CPS rely on accurate knowledge of operator positions, activities, and environmental conditions to support ergonomic task allocation, safe interaction with collaborative robots, and context-aware decision support \cite{reddy2025humanTrustCPS, kadam2025humanCentricCPPS}. RTLS technologies thus form an important bridge between physical industrial processes and the human-centred intelligence that characterizes the transition toward Industry~5.0 \cite{lachvajderova2024real, pulcini2023humanCentricRTLS, kadam2025humanCentricCPPS}.

\subsection{Structural Challenges in RTLS Deployment}
Despite their potential, RTLS deployments frequently encounter substantial technical, financial, and organizational challenges \cite{organiciak2025indoorRTLSmultistory, kuepper2022automotiveRTLS5G}. Many industrial facilities were not designed with RTLS in mind and exhibit structural features that complicate installation, such as high ceilings, limited mounting options, and extensive use of metallic structures and heavy machinery. These characteristics can lead to signal obstruction, severe multipath propagation, and coverage gaps \cite{grigorovich2021rtls, grigorovichFactorsAffectingImplementation2021, organisciak2025indoor}. Retrofitting facilities with additional cabling, power outlets, and network infrastructure further increases deployment cost and may disrupt operations \cite{cano2023iotbuilding, luder2025anitrack}.

Achieving and maintaining high localization accuracy typically requires careful anchor placement and calibration, often involving site surveys, RF modelling, and iterative tuning \cite{liu2023positioncalibration, volpiLowCostRealTimeLocating2023}. Environmental changes and hardware modifications can degrade performance over time, necessitating periodic recalibration \cite{zradzinskiEvaluationSARHuman2020}. These activities demand specialized expertise, which may be unavailable in-house in many organizations, particularly SMEs \cite{bendavidRisePassiveRFID2024, sharma2025predictability}.

RTLSs also raise privacy and cybersecurity concerns, as they generate detailed traces of asset and personnel movement \cite{bai2025federated}. Network vulnerabilities associated with Wi-Fi or Bluetooth protocols may expose RTLS infrastructures to unauthorized access or data interception, increasing the need for robust security measures such as end-to-end encryption, identity management, and continuous threat monitoring \cite{shaikhanova2025vulnerability, fomchenkova2022neurocrypto}. Finally, integration of RTLS data into existing ERP, WMS, and MES systems often requires custom middleware and data-engineering effort \cite{fettaPolicyAnalysisReturn2023, yosephine2025inventory}, contributing to total cost of ownership and extending deployment timelines.

\section{Operational Value and Limitations of RTLS}
\subsection{Operational Value in Manufacturing and Logistics}
RTLSs provide two distinct layers of value. At the measurement layer, they generate \emph{raw location data}: time-stamped position estimates for tagged assets and personnel within the monitored space. At the decision-support layer, this data is transformed into \emph{actionable intelligence} when processed by WMS, MES, analytics engines, or digital twins to identify bottlenecks, detect anomalies, trigger alarms, or optimize resource allocation \cite{nurdiyanto2024MESdigital, rodic2017simulationI40}.

In manufacturing and warehousing, RTLSs have been shown to improve asset and inventory management by reducing search times for tools, materials, and equipment, decreasing the incidence of misplaced inventory, and mitigating overstocking \cite{baviskar2025realtime, prauseChallengesIndustry402019}. Analyses of movement patterns support optimization of material flows, reduction of delays on assembly lines and in storage areas, and better synchronization with just-in-time supply chains \cite{rahayuCanggahCoffeeProcessed2021}. These improvements can translate into increased throughput, reduced work-in-process inventory, and lower operating costs.

In safety-critical environments, RTLSs enhance worker protection by monitoring personnel in hazardous areas, enforcing digital geofences, and supporting more effective emergency evacuations and muster procedures \cite{daher2024rtlsOilGas, haleem2024safety}. Integration with predictive-maintenance platforms enables continuous monitoring of equipment usage and environmental conditions, supporting proactive interventions before failures occur \cite{lukito2025predictive}. In sectors such as healthcare, pharmaceuticals, and food, RTLS-enabled traceability supports compliance with regulatory and quality requirements \cite{overmannRealtimeLocatingSystems2021, gladyszAPPROACHRTLSSELECTION2018}.

\subsection{Decision Support and Strategic Planning}
Beyond immediate operational benefits, RTLSs enable more sophisticated, data-driven management practices. Many organizations, particularly SMEs, lack dedicated analytics teams. RTLS data, when combined with business-intelligence tools, can reveal underutilized assets, systematically identify bottlenecks, and support scenario analysis for layout changes or process redesign \cite{guoCorrelationAnalysisInformatization2023}. When embedded in digital twins, RTLS data allows “what-if” simulations that inform strategic decisions in capacity planning, automation investments, and safety-system design \cite{aouani2023digitalTwinWarehouse}.

However, exploiting this strategic potential requires more than simply collecting location data. RTLS data must be integrated into existing planning and control processes, and appropriate metrics and visualization tools must be in place. Without adequate integration and interpretation, there is a risk that RTLSs add data volume without improving decision quality.

\subsection{Limitations and Hidden Costs}
The limitations of RTLSs are closely linked to their deployment model. Achieving high accuracy in complex industrial environments typically requires:
\begin{itemize}
    \item sufficiently dense anchor layouts to ensure line-of-sight coverage and geometric diversity;
    \item careful selection of mounting locations and heights;
    \item reliable networking and power at each anchor location; and
    \item time-consuming calibration and testing phases.
\end{itemize}
These requirements directly translate into monetary cost, engineering effort, and operational disruption \cite{vanherbruggen2024twr, mutzeUseCasesRealTime2021, tercas2024bayesian}. RTLS accuracy can be degraded by radio-frequency interference, extensive metallic structures, and dynamic obstructions such as forklifts or stacked materials \cite{bastiaens2024vlp, bendavidRisePassiveRFID2024}, necessitating conservative design margins or more frequent recalibration.

In addition, integration of RTLS data into legacy ERP, WMS, and MES systems is often non-trivial. Proprietary or poorly documented interfaces can require custom middleware, API development, or vendor-specific adapters \cite{yosephine2025inventory, fettaPolicyAnalysisReturn2023}. These “soft” costs are often under-emphasized in vendor documentation but loom large in real deployments and are especially problematic for SMEs with limited IT resources.

\section{RTLS Adoption Barriers in Small and Medium-Sized Enterprises}
\label{sec:rtls_installation_barrier}
\subsection{Economic and Infrastructure Constraints}
SMEs are central to industrial economies and play a key role in job creation and innovation \cite{kumar2025msme}. At the same time, they typically operate under tight capital constraints and face difficulties accessing external finance \cite{okonMSMEsEngineEconomic2018}. High upfront investment is therefore a primary barrier when considering RTLS deployments. Costs include hardware (tags, anchors, gateways), software licensing or development, installation, calibration, training, and ongoing maintenance \cite{runyonUnderstandingAthleticTrainers2020, yosephine2025iotinventory}. Unlike large enterprises, SMEs cannot easily amortize RTLS investments across multiple sites.

Uncertainty about return on investment (ROI) exacerbates the challenge. Quantifying the savings from improved inventory accuracy, reduced search times, or fewer stockouts is non-trivial, particularly when baseline processes are informal or weakly instrumented \cite{zahraBoostingEmergingTechnology2021}. As a result, even when managers recognize the conceptual value of RTLS, they may be unable to construct a sufficiently compelling business case to justify expenditure.

Infrastructure constraints further increase deployment difficulty. Many SMEs operate in older or leased facilities that were not designed with RTLS, or even pervasive networking, in mind. High ceilings, lack of suitable mounting points, and limited power and network outlets complicate anchor placement and wiring \cite{grigorovich2021rtls, cano2023iotbuilding}. Retrofitting such facilities with additional cabling and power distribution can be costly and disruptive \cite{luder2025anitrack}. Multi-floor layouts and dense shelving exacerbate non-line-of-sight propagation and multipath, requiring more anchors or careful RF modelling \cite{mutzeUseCasesRealTime2021, jesusazabal2025anchoropt}.

\subsection{Integration and Data-Management Challenges}
RTLS data delivers its greatest value when tightly integrated with existing ERP, WMS, or MES platforms. Many SMEs, however, rely on legacy systems or ad hoc combinations of spreadsheets and manual processes \cite{liEvaluationDigitalTransformation2022, ghobakhlooDriversBarriersIndustry2022}. When RTLS solutions expose proprietary or non-standard interfaces, integration projects can become complex and expensive, requiring custom middleware, data pipelines, and ongoing maintenance \cite{stentoftDriversBarriersIndustry2019, yosephine2025inventory}. For SMEs with limited IT staff, this represents a significant barrier.

From an SME perspective, there is a risk that RTLS becomes an isolated dashboard rather than a source of operational intelligence that directly influences planning, replenishment, and scheduling. The need for bespoke integration increases perceived project risk and extends time-to-value. Moreover, RTLSs introduce new data-governance and cybersecurity requirements. Detailed traces of asset and personnel movement can be sensitive, and network vulnerabilities associated with Wi-Fi or Bluetooth may expose RTLS infrastructures to unauthorized access or data interception \cite{bai2025federated, shaikhanova2025vulnerability, fomchenkova2022neurocrypto}. Designing and maintaining appropriate security measures is a non-trivial task for organizations with limited technical capacity.

\subsection{Human and Organizational Factors}
Human and organizational factors constitute another major adoption barrier. Employees may perceive RTLS as a surveillance tool rather than an operational aid, particularly when personnel tracking is involved \cite{kirRealTimeLocationSystem2023}. Concerns about monitoring and performance evaluation can generate resistance, especially if the benefits for workers are not clearly articulated. Effective change management and communication are therefore essential components of successful deployment.

SMEs rarely have dedicated IT or OT teams capable of managing complex RTLS dashboards, performing root-cause analysis when anomalies arise, or maintaining secure network configurations \cite{sharma2025predictability}. Training staff on RTLS usage requires time and resources and competes with day-to-day operational demands \cite{ullahAnalyzingBarriersImplementation2021}. Without user-friendly interfaces and targeted training, there is a risk that RTLS systems are underutilized or abandoned. These human factors interact with economic and infrastructure constraints to create a structural barrier to adoption, even in organizations that recognize the potential value of enhanced visibility.

\section{Survey Evidence on Practitioner Perceptions of RTLS}
To complement the literature analysis, an online survey of sixteen manufacturing and technology professionals across Canada and the United States was conducted. Respondents included chief technology officers, engineering leads, operations managers, and research scientists. Their organizations ranged from small production facilities to large manufacturing plants, with floor space spanning from less than \SI{5000}{\sqft} to more than \SI{25000}{\sqft}. The survey examined perceptions of RTLS value, cost, and infrastructure complexity, with a particular focus on adoption barriers in SME-like settings.

\subsection{Survey Design, Respondent Profile, and Inventory Challenges}
\begin{figure}[!ht]
\centering
\textbf{Respondents' Company Facility Sizes}\\[4pt]
\includegraphics[width=1\linewidth]{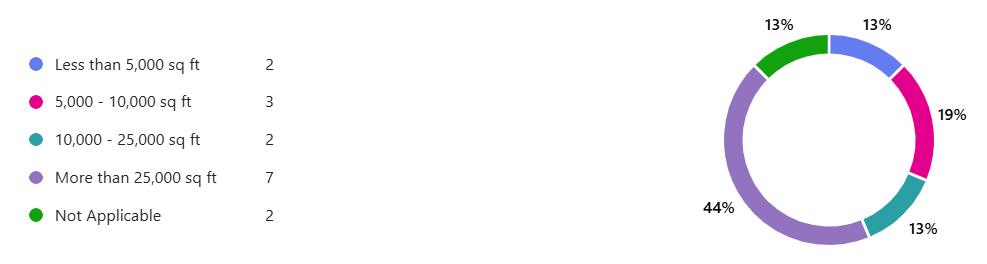}
\caption{Reported industrial facility sizes among survey respondents.}
\label{fig:Manufacturing Facility Size}
\end{figure}

Figure~\ref{fig:Manufacturing Facility Size} summarizes the reported facility sizes. Almost half of the respondents (\(7/16 \approx 44\%\)) worked in facilities larger than \SI{25000}{\sqft}, while smaller groups reported operating in facilities of less than \SI{5000}{\sqft} (2 responses), \SIrange{5000}{10000}{\sqft} (3 responses), and \SIrange{10000}{25000}{\sqft} (2 responses). Two respondents indicated that facility size was not directly applicable to their role.

\begin{figure}[htbp]
\centering
\textbf{Respondents' Opinion on the Most Challenging Aspects of Inventory Management}\\[4pt]
\includegraphics[width=\linewidth]{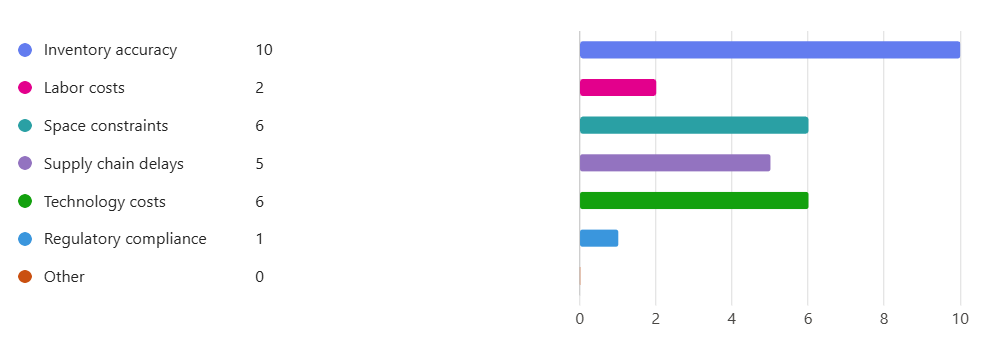}
\caption{Respondents' opinion on the most challenging aspects of inventory management.}
\label{fig:Inventory Mgt Challenges}
\end{figure}

Before addressing RTLS explicitly, the survey probed general inventory-management pain points (Figure~\ref{fig:Inventory Mgt Challenges}). Inventory accuracy emerged as the dominant challenge, selected by 10 out of 16 respondents. Technology costs and space constraints were each selected by 6 respondents, and supply-chain delays by 5 respondents. Labour costs and regulatory compliance were mentioned much less frequently. This pattern suggests that many organizations already experience operational friction that could, in principle, be alleviated by improved visibility and tracking.

\begin{figure}[ht]
\centering
\textbf{Respondents' Company Familiarity with RTLS Systems}\\[4pt]
\includegraphics[width=1\linewidth]{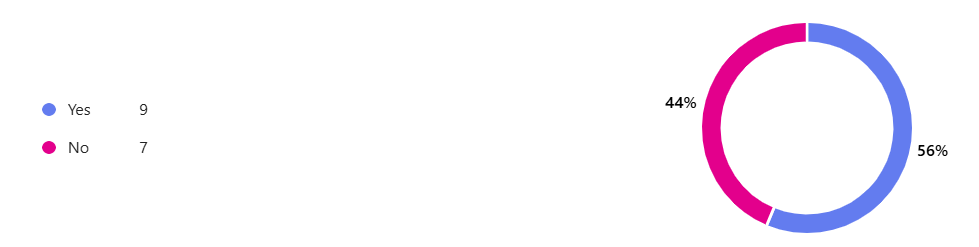}
\caption{More than half of respondents report some familiarity with RTLS systems for inventory management.}
\label{fig:Familiarity with RTLS for Inventory}
\end{figure}

RTLS awareness among respondents was moderate but not universal (Figure~\ref{fig:Familiarity with RTLS for Inventory}). A slight majority (\(9/16 \approx 56\%\)) reported that their company was familiar with RTLS systems for inventory management, while 7 respondents (44\%) indicated no prior familiarity. Lack of awareness therefore remains a barrier for a substantial subset of potential adopters.

\subsection{Perceived Importance, Cost Drivers, and Acceptable Investment Levels}
\begin{figure}[ht]
\centering
\textbf{Respondents' Opinion on the Importance of RTLS Systems}\\[4pt]
\includegraphics[width=1\linewidth]{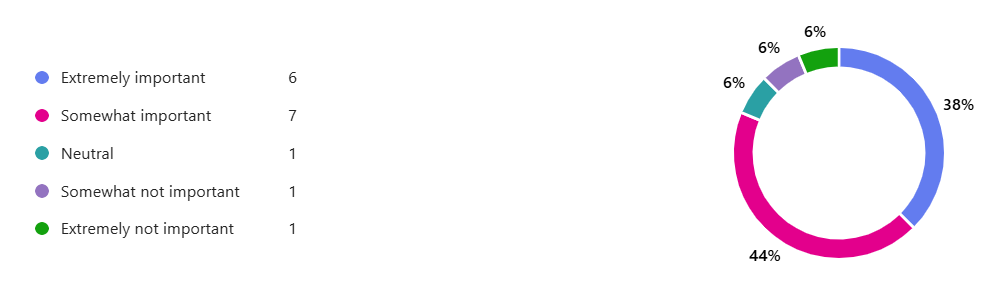}
\caption{A large majority rate real-time inventory tracking as somewhat or extremely important.}
\label{fig:RTLS Benefits}
\end{figure}

Among all respondents, the perceived importance of real-time inventory tracking was high (Figure~\ref{fig:RTLS Benefits}). Six respondents rated it as \emph{extremely important} and seven as \emph{somewhat important}, so \(13/16\) (approximately 81\%) viewed real-time tracking as important for operational efficiency. Only three respondents were neutral or negative. This confirms that practitioners generally recognize the potential benefits of RTLS, even if they have not yet adopted such systems.

\begin{figure}[ht]
\centering
\textbf{Respondents' Opinion on RTLS Cost of Implementation}\\[4pt]
\includegraphics[width=1\linewidth]{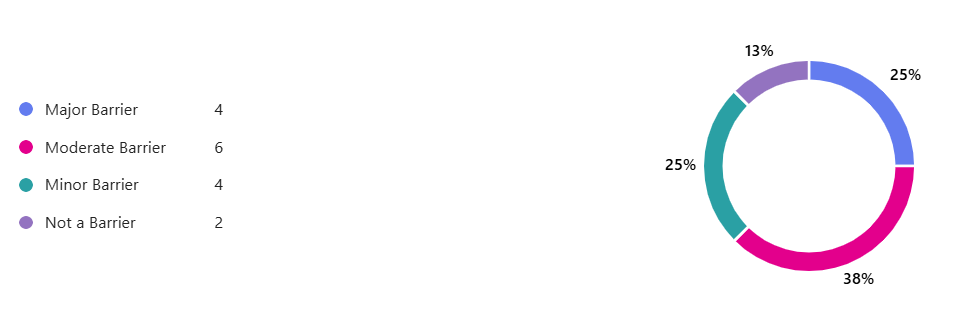}
\caption{Most respondents regard RTLS implementation cost as a moderate or major barrier.}
\label{fig:Cost of Implementation Barrier}
\end{figure}

When asked directly about the cost of implementation (Figure~\ref{fig:Cost of Implementation Barrier}), around two-thirds of respondents rated cost as a \emph{moderate} or \emph{major} barrier, with only a small minority viewing it as a minor or negligible concern. This reinforces the centrality of cost in adoption decisions, particularly in SME-like environments.

\begin{figure}[ht]
\centering
\textbf{Respondents' Opinion on Highest Infrastructure Cost Drivers}\\[4pt]
\includegraphics[width=1\linewidth]{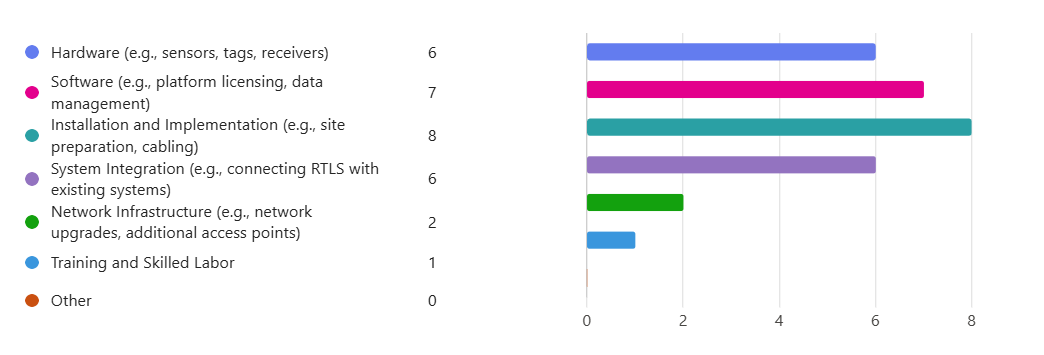}
\caption{Installation/implementation and software licensing emerge as the dominant perceived cost drivers, with hardware and integration also significant.}
\label{fig:Highest Infrastructure Cost Driver}
\end{figure}

To better understand where these costs arise, the survey asked respondents to identify the most significant infrastructure cost drivers for an RTLS deployment (Figure~\ref{fig:Highest Infrastructure Cost Driver}). Installation and implementation (including site preparation, cabling, and physical deployment) was selected by 8 respondents, making it the most frequently cited cost driver. Software and data-management platforms were next (7 responses), followed by hardware components (tags, sensors, receivers) and system integration, each selected by 6 respondents. Network upgrades and training were mentioned less often. Practitioners thus view the “soft” costs associated with installing, configuring, and integrating RTLS infrastructure as at least as important as the hardware bill of materials.

\begin{figure}[htbp]
\centering
\textbf{Respondents' Ratings of Key RTLS Adoption Factors}\\[4pt]
\includegraphics[width=\linewidth]{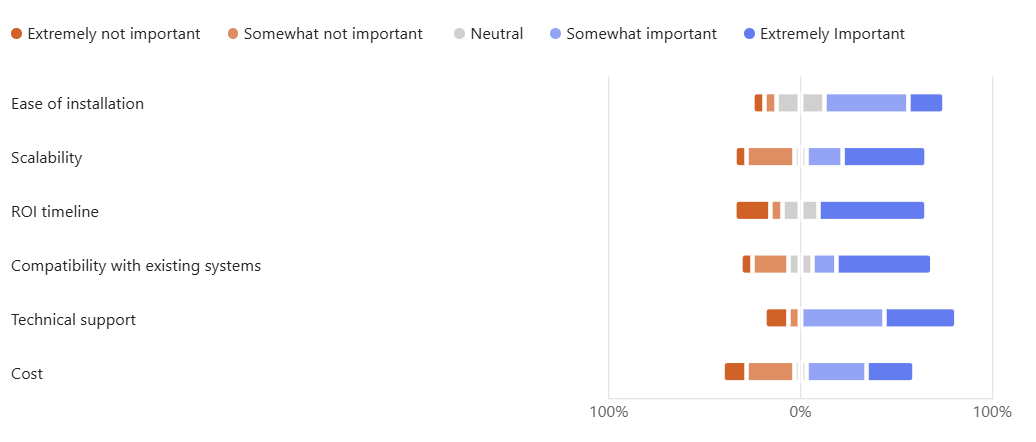}
\caption{Ease of installation, cost, ROI timeline, and compatibility with existing systems are rated as the most critical factors influencing RTLS adoption.}
\label{fig:Factors Influencing RTLS Adoption}
\end{figure}

A separate question explored which factors are most influential in RTLS adoption decisions (Figure~\ref{fig:Factors Influencing RTLS Adoption}). Across criteria including ease of installation, scalability, ROI timeline, compatibility with existing systems, technical support, and cost, the majority of responses clustered in the \emph{somewhat important} and \emph{extremely important} categories. Qualitatively, ease of installation, cost, ROI timeline, and compatibility attracted the highest proportion of \emph{extremely important} ratings, indicating a preference for solutions that can be deployed quickly, integrate cleanly with current systems, and deliver a clear payback.

\begin{figure}[ht]
\centering
\textbf{Respondents' Opinion on Reasonable RTLS Implementation Costs}\\[4pt]
\includegraphics[width=1\linewidth]{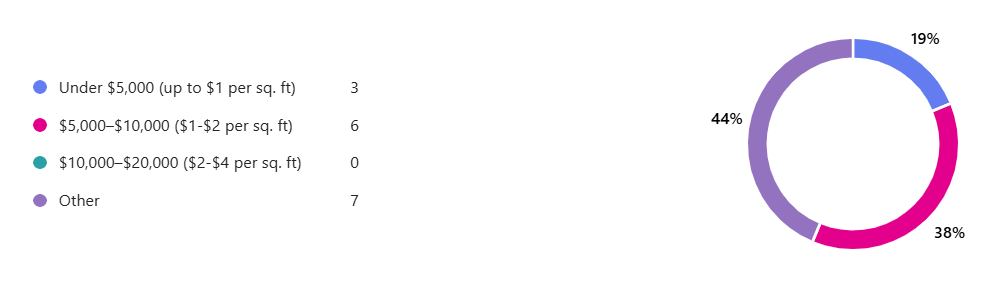}
\caption{More than half of respondents specify an acceptable upfront RTLS investment below \$10{,}000; none select \$10{,}000--\$20{,}000 as a preferred range.}
\label{fig:Reasonable Upfront Infrastructure Cost}
\end{figure}

Respondents were also asked what level of upfront investment would be considered reasonable for an RTLS deployment in their context (Figure~\ref{fig:Reasonable Upfront Infrastructure Cost}). Three respondents selected ``under \$5{,}000'' and six selected ``\$5{,}000--\$10{,}000,'' corresponding to approximately \$1–\$2 per square foot for typical SME facilities. No respondent chose the higher ``\$10{,}000--\$20{,}000'' bracket, while seven selected ``Other,'' often noting that acceptable cost would depend on facility size, features, or demonstrable ROI. Overall, roughly 56\% of respondents explicitly indicated an upper bound below \$10{,}000, suggesting relatively tight capital-expenditure constraints for RTLS projects.

\subsection{Infrastructure Expectations and Anchor-Node Deployment}
\begin{figure}[ht]
\centering
\textbf{Respondents' Opinion on RTLS Node Minimization vs.\ System Adoption}\\[4pt]
\includegraphics[width=1\linewidth]{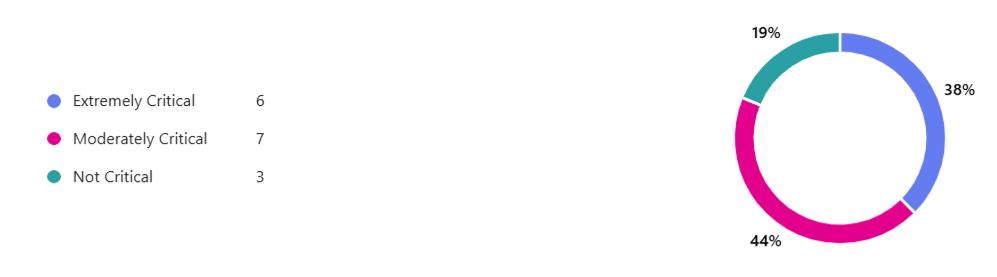}
\caption{More than 80\% of respondents consider minimizing the number of anchor nodes to be moderately or extremely critical to RTLS adoption.}
\label{fig:RTLS Node Minimization vs Adoption}
\end{figure}

Given the focus of this work on low-infrastructure architectures, several survey items probed attitudes toward anchor-node deployment. Respondents were first asked to rate how critical it would be to minimize the number of anchor nodes when deciding whether to adopt an RTLS (Figure~\ref{fig:RTLS Node Minimization vs Adoption}). Six respondents (38\%) rated anchor minimization as \emph{extremely critical}, seven (44\%) as \emph{moderately critical}, and only three (19\%) as \emph{not critical}. Thus, \(13/16\) respondents (approximately 81\%) viewed a low anchor count as at least moderately important to their adoption decision.

\begin{figure}[ht]
\centering
\textbf{Respondents' Opinion on RTLS Five-Node Surface Area Coverage}\\[4pt]
\includegraphics[width=1\linewidth]{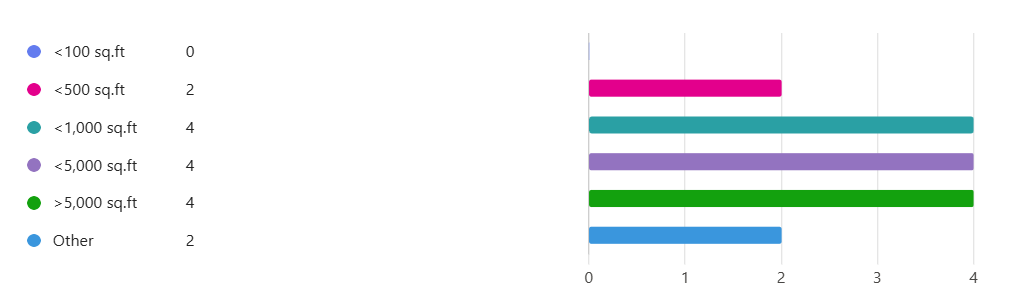}
\caption{Respondents exhibit diverse beliefs about the surface area that five RTLS nodes can cover, reflecting uncertainty about system scalability and coverage efficiency.}
\label{fig:Maximum Coverage area of 5 nodes}
\end{figure}

The survey also asked respondents to estimate the surface area that five anchor nodes could reasonably cover in a typical industrial environment (Figure~\ref{fig:Maximum Coverage area of 5 nodes}). Responses were widely distributed: two respondents selected ``less than \SI{500}{\sqft},'' four selected ``less than \SI{1000}{\sqft},'' another four selected ``less than \SI{5000}{\sqft},'' four selected ``greater than \SI{5000}{\sqft},'' and two chose ``Other.'' Notably, no respondents selected the smallest category (``less than \SI{100}{\sqft}''). This dispersion reflects substantial uncertainty and a lack of shared expectations regarding realistic coverage, highlighting the need for clearer benchmarks and more intuitive deployment models for SMEs.

\begin{figure}[ht]
\centering
\textbf{Respondents' Opinion on RTLS Adoption Concerns Besides Cost}\\[4pt]
\includegraphics[width=1\linewidth]{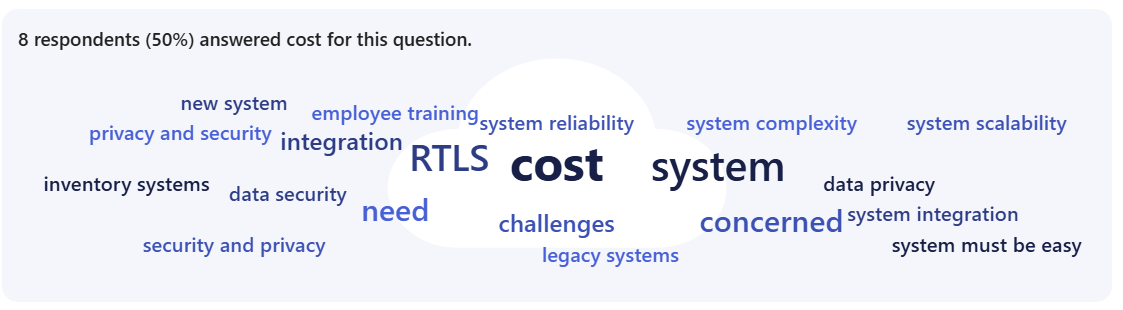}
\caption{Summary of open-ended responses indicating additional concerns about RTLS adoption beyond cost, including integration complexity and long-term reliability.}
\label{fig:Concern Beside Cost for RTLS Adoption}
\end{figure}

Open-ended questions (summarized in Figure~\ref{fig:Concern Beside Cost for RTLS Adoption}) highlighted recurring concerns beyond cost, including:
\begin{itemize}
    \item difficulties integrating RTLS data with existing ERP, WMS, or MES platforms;
    \item uncertainty about how to quantify ROI and payback periods;
    \item doubts about long-term reliability and vendor support; and
    \item a shortage of in-house technical expertise for installation, calibration, and maintenance.
\end{itemize}

Table~\ref{tab:survey_summary} collates several key quantitative indicators from the survey.

\begin{table}[ht]
\centering
\caption{Summary of Selected Survey Indicators on RTLS Adoption}
\begin{tabular}{p{4cm} p{3cm} p{5cm}}
\toprule
\textbf{Metric} & \textbf{Finding} & \textbf{Interpretation} \\
\midrule
RTLS Familiarity & 56\% & Moderate awareness; nearly half still unfamiliar \\
Rated RTLS Important & 81\% & Operational value widely recognized \\
Moderate/Major Cost Barrier & $\sim$65\% & Cost remains a dominant constraint \\
Acceptable Cost Range & \$5{,}000--\$10{,}000 & Typical SME affordability limit for upfront investment \\
Anchor Minimization Criticality & 81\% (Moderate + High) & Strong preference for low-infrastructure setups \\
\bottomrule
\end{tabular}
\label{tab:survey_summary}
\end{table}

Overall, the survey evidence is consistent with the qualitative analysis in Sections~3 and~4: practitioners generally perceive RTLS as operationally valuable, but regard implementation cost and infrastructure complexity, particularly installation, integration, and calibration effort, as substantial barriers to adoption. The strong emphasis on ease of installation, low infrastructure dependency, and limited anchor counts provides empirical motivation for developing simplified, anchor-minimizing or anchor-free localization frameworks, such as those proposed in emerging models like Protraction Theory.

\section{Innovation Opportunities and Design Implications for SME-Oriented RTLS}
\subsection{Simplified Deployment and Cost Models}
The barriers identified in Sections~\ref{sec:rtls_installation_barrier} and~5 do not suggest that RTLSs are intrinsically unsuitable for SMEs; rather, they indicate that prevailing deployment models are misaligned with SME constraints. A central design priority is therefore to streamline installation and reduce capital expenditure. Wireless RTLS architectures based on BLE, UWB, or battery-powered RFID can eliminate portions of the cabling and physical infrastructure that currently drive installation cost \cite{chakrabortyExploratorySurveyAustralian2023}. Integration of energy-harvesting mechanisms or ultra-low-power communication schemes can further reduce battery replacement frequency and maintenance overhead \cite{cansizRadioFrequencyEnergy2019, velooHybridSolarRFEnergy2023}.

Modular deployment strategies allow SMEs to start with a limited set of high-value use cases (for example, tracking critical tools in maintenance areas or specific high-value inventory zones) and expand coverage over time. RTLS-as-a-Service (RTLSaaS) models, in which hardware and software are bundled into subscription offerings, can transform large upfront capital costs into more manageable operating expenses, thereby reducing financial barriers to initial experimentation. Cloud-native platforms further reduce the need for on-premise server infrastructure and support remote management, over-the-air updates, and outsourced data security \cite{bakytbekov3DPrintedBifunctional2020}.

\subsection{Self-Calibration, Intelligence, and User-Centred Analytics}
Traditional RTLS deployments often require manual calibration, including distance measurements, environment mapping, and parameter tuning. Advances in AI-driven self-calibration algorithms offer an opportunity to reduce this burden. RTLS systems can leverage machine-learning models to adjust anchor parameters, signal-strength thresholds, and localization models automatically based on observed data \cite{ngContactTracingUsing2022}. Over time, the system can adapt to environmental changes and interference patterns, reducing the need for manual intervention and periodic full recalibration.

Given that many SMEs do not have dedicated analytics teams, modern RTLS solutions should also emphasize user-centred design. Dashboards that provide intuitive heat maps, simple asset search, and actionable alerts can make location data accessible to non-experts \cite{giovandoEvolutionsManufacturingCost2020}. Embedding basic analytics, such as dwell-time analysis, bottleneck identification, and simple what-if comparisons, reduces dependence on separate business-intelligence tools. Automated reporting (for instance, weekly summaries of lost time due to searching for assets, or heat maps of congestion zones) can help SMEs build an internal business case for scaling RTLS beyond initial pilots while supporting continuous-improvement initiatives.

\subsection{Low-Infrastructure and Secure RTLS Architectures}
The survey results demonstrate a strong practitioner preference for minimizing the number of anchor nodes. This preference aligns with ongoing research into anchor-minimizing and anchor-free localization methods, which seek to reduce the infrastructure footprint required for acceptable accuracy. Such approaches may leverage:
\begin{itemize}
    \item hybrid localization that combines radio-based ranging with inertial sensors and environment constraints;
    \item opportunistic use of existing infrastructure (such as Wi-Fi access points, cellular base stations, or other deployed beacons);
    \item cooperative localization between tags; and
    \item new theoretical frameworks, such as Protraction Theory, that aim to infer relative positions with fewer fixed references.
\end{itemize}
Although many of these methods are still under active development, they offer promising avenues for designing RTLS solutions that are structurally better aligned with SME resource constraints.

As RTLS data becomes more integral to operations and supply chains, secure and trusted data sharing is also critical. Integrating RTLS with blockchain or distributed-ledger technologies has been proposed as a means to ensure tamper-proof traceability in high-compliance sectors \cite{hsiaoOptimizationBasedApproachesMinimizing2021, tijan2019blockchainLogistics}. For SMEs, however, practical implementations must balance security with simplicity; managed security services and lightweight identity-management schemes may be more appropriate than fully decentralized architectures. Standardized, well-documented APIs (e.g., REST, MQTT, or OPC-UA) and reference integration patterns \cite{katariaRealTimeLocationSystems2024, stentoftDriversBarriersIndustry2019} can further lower integration effort and reduce vendor lock-in.

\section{Conclusion}
RTLSs have emerged as a central enabling technology for Industry~4.0, Logistics~4.0, and the transition toward human-centric Industry~5.0. They provide the spatial and temporal visibility required to optimize workflows, enhance safety, and underpin digital-twin and AI-driven decision-support systems. However, current RTLS deployment models, characterized by dense anchor infrastructures, extensive cabling, bespoke integration, and manual calibration, are often misaligned with the economic, infrastructural, and organizational constraints faced by SMEs.

This paper combined a structured review of RTLS architectures and deployment challenges with survey evidence from sixteen manufacturing and technology professionals. The analysis indicates that practitioners generally regard real-time tracking as operationally important, but perceive implementation cost and complexity, driven largely by installation, integration, and calibration effort, as major barriers to adoption. Many respondents reported acceptable upfront investment levels below \$10{,}000 and expressed a strong preference for low-infrastructure architectures with minimized anchor counts.

These results suggest that RTLS adoption in SMEs is constrained less by a lack of perceived value than by a misfit between prevailing RTLS offerings and SME requirements. Addressing this misfit will require innovation along several dimensions, including wireless and modular architectures, self-calibrating and AI-augmented localization, user-centred analytics, standardized integration interfaces, and anchor-minimizing or anchor-free localization frameworks such as Protraction Theory. Future work should extend the empirical evidence base through larger surveys and field trials in diverse SME environments, quantify the impact of alternative deployment models on total cost of ownership, and evaluate emerging low-infrastructure localization methods in realistic industrial conditions. Aligning RTLS design with the realities of SME operations is essential for ensuring that the benefits of Industry~4.0 and Industry~5.0 are accessible across the full spectrum of industrial firms.

\section*{Acknowledgments}
\addcontentsline{toc}{section}{Acknowledgments}

The authors gratefully acknowledge the support of Eris Canada Solutions Inc.\ (\url{https://eriscanada.com}) for facilitating industry contacts and providing valuable industrial perspectives that informed this work. We also acknowledge financial support from the Vadasz Scholars Program at McGill University (\url{https://www.mcgill.ca/vadaszscholars/}). The authors thank colleagues in the McGill Networks Research Lab for their feedback and stimulating discussions, and express their gratitude to friends and family for their continuous encouragement and support.

\bibliographystyle{IEEEtran}
\bibliography{White5Paper}

@inproceedings{bakytbekov3DPrintedBifunctional2020,
  title = {{{3D Printed Bifunctional Triple-Band Heatsink Antenna}} for {{RF}} and {{Thermal Energy Harvesting}}},
  booktitle = {2020 {{IEEE International Symposium}} on {{Antennas}} and {{Propagation}} and {{North American Radio Science Meeting}}},
  author = {Bakytbekov, Azamat and Iman, Zere and Shamim, Atif},
  year = {2020},
  month = jul,
  pages = {1563--1564},
  publisher = {IEEE},
  address = {Montreal, QC, Canada},
  doi = {10.1109/IEEECONF35879.2020.9330253},
  urldate = {2025-03-17},
  copyright = {https://ieeexplore.ieee.org/Xplorehelp/downloads/license-information/IEEE.html},
  isbn = {978-1-7281-6670-4}
}

@article{bendavidRisePassiveRFID2024,
  title = {The {{Rise}} of {{Passive RFID RTLS Solutions}} in {{Industry}} 5.0},
  author = {Bendavid, Ygal and Rostampour, Samad and Berrabah, Yacine and Bagheri, Nasour and Safkhani, Masoumeh},
  year = {2024},
  month = mar,
  journal = {Sensors},
  volume = {24},
  number = {5},
  pages = {1711},
  issn = {1424-8220},
  doi = {10.3390/s24051711},
  urldate = {2025-03-17},
  copyright = {https://creativecommons.org/licenses/by/4.0/},
  langid = {english}
}

@article{cansizRadioFrequencyEnergy2019,
  title = {Radio {{Frequency Energy Harvesting}} with {{Different Antennas}} and {{Output Powers}}},
  author = {Cansiz, Mustafa},
  year = {2019},
  month = jul,
  journal = {Balkan Journal of Electrical and Computer Engineering},
  volume = {7},
  number = {3},
  pages = {245--249},
  issn = {2147-284X},
  doi = {10.17694/bajece.551790},
  urldate = {2025-03-17}
}

@article{chakrabortyExploratorySurveyAustralian2023,
  title = {Exploratory {{Survey}} of {{Australian SMEs}}: An {{Investigation}} into the {{Barriers}} and {{Opportunities Associated}} with {{Circular Economy}}},
  shorttitle = {Exploratory {{Survey}} of {{Australian SMEs}}},
  author = {Chakraborty, Ayon and Barton, Andrew and O'Loughlin, Andrew and Kandra, Harpreet S.},
  year = {2023},
  month = sep,
  journal = {Circular Economy and Sustainability},
  volume = {3},
  number = {3},
  pages = {1275--1297},
  issn = {2730-597X, 2730-5988},
  doi = {10.1007/s43615-022-00235-0},
  urldate = {2025-03-17},
  langid = {english}
}

@article{fettaPolicyAnalysisReturn2023,
  title = {Policy {{Analysis}} of {{Return}} to {{Learn After Sport}} and {{Recreational Related Concussion}} for {{Secondary Schools}} in {{New England}}: {{Relevance}} to {{School Nurses}} and {{Nursing Practice}}},
  shorttitle = {Policy {{Analysis}} of {{Return}} to {{Learn After Sport}} and {{Recreational Related Concussion}} for {{Secondary Schools}} in {{New England}}},
  author = {Fetta, Joseph M. and Starkweather, Angela R. and Van Hoof, Thomas and Huggins, Robert and Casa, Douglas and Gill, Jessica},
  year = {2023},
  month = nov,
  journal = {Policy, Politics, \& Nursing Practice},
  volume = {24},
  number = {4},
  pages = {278--287},
  issn = {1527-1544, 1552-7468},
  doi = {10.1177/15271544231186359},
  urldate = {2025-03-17},
  langid = {english}
}

@article{ghobakhlooDriversBarriersIndustry2022,
  title = {Drivers and Barriers of {{Industry}} 4.0 Technology Adoption among Manufacturing {{SMEs}}: A Systematic Review and Transformation Roadmap},
  shorttitle = {Drivers and Barriers of {{Industry}} 4.0 Technology Adoption among Manufacturing {{SMEs}}},
  author = {Ghobakhloo, Morteza and Iranmanesh, Mohammad and Vilkas, Mantas and Grybauskas, Andrius and Amran, Azlan},
  year = {2022},
  month = sep,
  journal = {Journal of Manufacturing Technology Management},
  volume = {33},
  number = {6},
  pages = {1029--1058},
  issn = {1741-038X},
  doi = {10.1108/JMTM-12-2021-0505},
  urldate = {2025-03-17},
  copyright = {https://www.emerald.com/insight/site-policies},
  langid = {english}
}

@article{giovandoEvolutionsManufacturingCost2020,
  title = {Evolutions in Manufacturing Cost Deployment},
  author = {Giovando, Guido and Crovini, Chiara and Venturini, Stefano},
  year = {2020},
  journal = {Global Business and Economics Review},
  volume = {22},
  number = {1/2},
  pages = {41},
  issn = {1097-4954, 1745-1329},
  doi = {10.1504/GBER.2020.105029},
  urldate = {2025-03-17},
  langid = {english}
}

@article{gladyszAPPROACHRTLSSELECTION2018,
  title = {{{AN APPROACH TO RTLS SELECTION}}},
  author = {Gladysz, B. and Santarek, K.},
  year = {2018},
  month = mar,
  journal = {DEStech Transactions on Engineering and Technology Research},
  number = {icpr},
  issn = {2475-885X},
  doi = {10.12783/dtetr/icpr2017/17576},
  urldate = {2025-03-17}
}

@article{grigorovichFactorsAffectingImplementation2021,
  title = {Factors {{Affecting}} the {{Implementation}}, {{Use}}, and {{Adoption}} of {{Real-Time Location System Technology}} for {{Persons Living With Cognitive Disabilities}} in {{Long-term Care Homes}}: {{Systematic Review}}},
  shorttitle = {Factors {{Affecting}} the {{Implementation}}, {{Use}}, and {{Adoption}} of {{Real-Time Location System Technology}} for {{Persons Living With Cognitive Disabilities}} in {{Long-term Care Homes}}},
  author = {Grigorovich, Alisa and Kulandaivelu, Yalinie and Newman, Kristine and Bianchi, Andria and Khan, Shehroz S and Iaboni, Andrea and McMurray, Josephine},
  year = {2021},
  month = jan,
  journal = {Journal of Medical Internet Research},
  volume = {23},
  number = {1},
  pages = {e22831},
  issn = {1438-8871},
  doi = {10.2196/22831},
  urldate = {2025-03-17},
  langid = {english}
}

@inproceedings{guoCorrelationAnalysisInformatization2023,
  title = {The {{Correlation Analysis Between}} the {{Informatization Level}} and the {{Profit Growth Rate}} of {{Small}} and {{Medi-um-Sized Enterprises}}},
  booktitle = {Proceedings of the 2nd {{International Conference}} on {{Big Data Economy}} and {{Digital Management}}, {{BDEDM}} 2023, {{January}} 6-8, 2023, {{Changsha}}, {{China}}},
  author = {Guo, Xingcheng},
  year = {2023},
  publisher = {EAI},
  address = {Changsha, People's Republic of China},
  doi = {10.4108/eai.6-1-2023.2330352},
  urldate = {2025-03-17},
  isbn = {978-1-63190-402-8},
  langid = {english}
}

@article{hsiaoOptimizationBasedApproachesMinimizing2021,
  title = {Optimization-{{Based Approaches}} for {{Minimizing Deployment Costs}} for {{Wireless Sensor Networks}} with {{Bounded Estimation Errors}}},
  author = {Hsiao, Chiu-Han and Lin, Frank Yeong-Sung and Yang, Hao-Jyun and Huang, Yennun and Chen, Yu-Fang and Tu, Ching-Wen and Zhang, Si-Yao},
  year = {2021},
  month = oct,
  journal = {Sensors},
  volume = {21},
  number = {21},
  pages = {7121},
  issn = {1424-8220},
  doi = {10.3390/s21217121},
  urldate = {2025-03-17},
  copyright = {https://creativecommons.org/licenses/by/4.0/},
  langid = {english}
}

@article{jangAreSmartManufacturing2022,
  title = {Are Smart Manufacturing Systems Beneficial for All {{SMEs}}? {{Evidence}} from {{Korea}}},
  shorttitle = {Are Smart Manufacturing Systems Beneficial for All {{SMEs}}?},
  author = {Jang, Soojeen and Chung, Yanghon and Son, Hosung},
  year = {2022},
  month = jul,
  journal = {Management Decision},
  volume = {60},
  number = {6},
  pages = {1719--1743},
  issn = {0025-1747},
  doi = {10.1108/MD-12-2020-1632},
  urldate = {2025-03-17},
  copyright = {https://www.emerald.com/insight/site-policies},
  langid = {english}
}

@article{katariaRealTimeLocationSystems2024,
  title = {Real-{{Time Location Systems}}: {{Revolutionizing Porter Services}} for {{Improving Patient Care Efficiency}}},
  shorttitle = {Real-{{Time Location Systems}}},
  author = {Kataria, Samriddhi and Sharma, Pankaj and Kataria, Suryansh and Saharan, Meenu and Sikka, Rajiv},
  year = {2024},
  month = sep,
  journal = {Cureus},
  issn = {2168-8184},
  doi = {10.7759/cureus.69922},
  urldate = {2025-03-17},
  langid = {english}
}

@article{kirRealTimeLocationSystem2023,
  title = {A {{Real-Time Location System Design}} for {{Production Facilities Working}} under {{COVID-19 Pandemic Precautions}}},
  author = {Kir, Sena},
  year = {2023},
  month = mar,
  journal = {Journal of Intelligent Systems: Theory and Applications},
  volume = {6},
  number = {1},
  pages = {34--42},
  issn = {2651-3927},
  doi = {10.38016/jista.1015515},
  urldate = {2025-03-17},
  langid = {english}
}

@article{liEvaluationDigitalTransformation2022,
  title = {Evaluation of {{Digital Transformation Maturity}} of {{Small}} and {{Medium-Sized Entrepreneurial Enterprises Based}} on {{Multicriteria Framework}}},
  author = {Li, Lan},
  editor = {Yang, Zaoli},
  year = {2022},
  month = jul,
  journal = {Mathematical Problems in Engineering},
  volume = {2022},
  pages = {1--11},
  issn = {1563-5147, 1024-123X},
  doi = {10.1155/2022/7085322},
  urldate = {2025-03-17},
  copyright = {https://creativecommons.org/licenses/by/4.0/},
  langid = {english}
}

@article{mutzeUseCasesRealTime2021,
  title = {Use {{Cases}} of {{Real-Time Locating Systems}} for {{Factory Planning}} and {{Production Monitoring}}},
  author = {M{\"u}tze, Alexander and Hingst, Lennart and Rochow, Niklas Eduard and Miebach, Timo and Nyhuis, Peter},
  year = {2021},
  journal = {SSRN Electronic Journal},
  issn = {1556-5068},
  doi = {10.2139/ssrn.3857878},
  urldate = {2025-03-17},
  langid = {english}
}

@article{ngContactTracingUsing2022,
  title = {Contact Tracing Using Real-Time Location System ({{RTLS}}): A Simulation Exercise in a Tertiary Hospital in {{Singapore}}},
  shorttitle = {Contact Tracing Using Real-Time Location System ({{RTLS}})},
  author = {Ng, Guan Yee and Ong, Biauw Chi},
  year = {2022},
  month = oct,
  journal = {BMJ Open},
  volume = {12},
  number = {10},
  pages = {e057522},
  issn = {2044-6055, 2044-6055},
  doi = {10.1136/bmjopen-2021-057522},
  urldate = {2025-03-17},
  langid = {english},
}

@article{okonMSMEsEngineEconomic2018,
  title = {{{MSMEs}} as {{Engine}} of {{Economic Growth}} in {{Nigeria}}: {{Challenges}} and {{Prospects}} of {{Scalability}}},
  shorttitle = {{{MSMEs}} as {{Engine}} of {{Economic Growth}} in {{Nigeria}}},
  author = {Okon, Emmanuel Okokondem},
  year = {2018},
  month = jan,
  journal = {Australian Finance \& Banking Review},
  volume = {2},
  number = {1},
  pages = {1--10},
  issn = {2576-120X, 2576-1196},
  doi = {10.46281/afbr.v2i1.75},
  urldate = {2025-03-17},
  copyright = {http://creativecommons.org/licenses/by/4.0}
}

@article{overmannRealtimeLocatingSystems2021,
  title = {Real-Time Locating Systems to Improve Healthcare Delivery: {{A}} Systematic Review},
  shorttitle = {Real-Time Locating Systems to Improve Healthcare Delivery},
  author = {Overmann, Kevin M and Wu, Danny T.Y and Xu, Catherine T and Bindhu, Shwetha S and Barrick, Lindsey},
  year = {2021},
  month = jun,
  journal = {Journal of the American Medical Informatics Association},
  volume = {28},
  number = {6},
  pages = {1308--1317},
  issn = {1527-974X},
  doi = {10.1093/jamia/ocab026},
  urldate = {2025-03-17},
  copyright = {https://academic.oup.com/journals/pages/open\_access/funder\_policies/chorus/standard\_publication\_model},
  langid = {english}
}

@article{prauseChallengesIndustry402019,
  title = {Challenges of {{Industry}} 4.0 {{Technology Adoption}} for {{SMEs}}: {{The Case}} of {{Japan}}},
  shorttitle = {Challenges of {{Industry}} 4.0 {{Technology Adoption}} for {{SMEs}}},
  author = {Prause, Martin},
  year = {2019},
  month = oct,
  journal = {Sustainability},
  volume = {11},
  number = {20},
  pages = {5807},
  issn = {2071-1050},
  doi = {10.3390/su11205807},
  urldate = {2025-03-17},
  copyright = {https://creativecommons.org/licenses/by/4.0/},
  langid = {english}
}

@article{rahayuCanggahCoffeeProcessed2021,
  title = {Canggah {{Coffee}} as the {{Processed Products}} of {{Micro Enterprises}}: {{System Design}} of {{E-Commerce}}},
  shorttitle = {Canggah {{Coffee}} as the {{Processed Products}} of {{Micro Enterprises}}},
  author = {Rahayu, Slamet and Iqbal, Muhammad and Ferdian, Nurizzi Rifqi and Fathurahman, Ferdi},
  year = {2021},
  month = mar,
  journal = {International Journal of Social Science and Business},
  volume = {5},
  number = {1},
  issn = {2549-6409, 2614-6533},
  doi = {10.23887/ijssb.v5i1.30755},
  urldate = {2025-03-17},
  copyright = {http://creativecommons.org/licenses/by-sa/4.0}
}

@article{runyonUnderstandingAthleticTrainers2020,
  title = {Understanding the {{Athletic Trainer}}'s {{Role}} in the {{Return-to-Learn Process}} at {{National Collegiate Athletic Association Division II}} and {{III Institutions}}},
  author = {Runyon, Lacey M. and Welch Bacon, Cailee E. and Neil, Elizabeth R. and Eberman, Lindsey E.},
  year = {2020},
  month = apr,
  journal = {Journal of Athletic Training},
  volume = {55},
  number = {4},
  pages = {365--375},
  issn = {1062-6050},
  doi = {10.4085/1062-6050-116-19},
  urldate = {2025-03-17},
  langid = {english}
}

@article{siagianSupplyChainIntegration2021,
  title = {Supply {{Chain Integration Enables Resilience}}, {{Flexibility}}, and {{Innovation}} to {{Improve Business Performance}} in {{COVID-19 Era}}},
  author = {Siagian, Hotlan and Tarigan, Zeplin Jiwa Husada and Jie, Ferry},
  year = {2021},
  month = apr,
  journal = {Sustainability},
  volume = {13},
  number = {9},
  pages = {4669},
  issn = {2071-1050},
  doi = {10.3390/su13094669},
  urldate = {2025-03-17},
  copyright = {https://creativecommons.org/licenses/by/4.0/},
  langid = {english}
}

@inproceedings{stentoftDriversBarriersIndustry2019,
  title = {Drivers and {{Barriers}} for {{Industry}} 4.0 {{Readiness}} and {{Practice}}: {{A SME Perspective}} with {{Empirical Evidence}}},
  shorttitle = {Drivers and {{Barriers}} for {{Industry}} 4.0 {{Readiness}} and {{Practice}}},
  booktitle = {Hawaii {{International Conference}} on {{System Sciences}}},
  author = {Stentoft, Jan and Jensen, Kent Wickstr{\o}m and Philipsen, Kristian and Haug, Anders},
  year = {2019},
  doi = {10.24251/HICSS.2019.619},
  urldate = {2025-03-17}
}

@article{szaszIndustry40Review2020,
  title = {Industry 4.0: A Review and Analysis of Contingency and Performance Effects},
  shorttitle = {Industry 4.0},
  author = {Sz{\'a}sz, Levente and Demeter, Krisztina and R{\'a}cz, B{\'e}la-Gergely and Losonci, D{\'a}vid},
  year = {2020},
  month = may,
  journal = {Journal of Manufacturing Technology Management},
  volume = {32},
  number = {3},
  pages = {667--694},
  issn = {1741-038X},
  doi = {10.1108/JMTM-10-2019-0371},
  urldate = {2025-03-17},
  copyright = {https://www.emerald.com/insight/site-policies},
  langid = {english}
}

@article{ullahAnalyzingBarriersImplementation2021,
  title = {Analyzing the Barriers to Implementation of Mass Customization in {{Indian SMEs}} Using Integrated {{ISM-MICMAC}} and {{SEM}}},
  author = {Ullah, Inayat and Narain, Rakesh},
  year = {2021},
  month = apr,
  journal = {Journal of Advances in Management Research},
  volume = {18},
  number = {2},
  pages = {323--349},
  issn = {0972-7981},
  doi = {10.1108/JAMR-04-2020-0048},
  urldate = {2025-03-17},
  copyright = {https://www.emerald.com/insight/site-policies},
  langid = {english}
}

@article{velooHybridSolarRFEnergy2023,
  title = {A {{Hybrid Solar-RF Energy Harvesting System Based}} on an {{EM4325-Embedded RFID Tag}}},
  author = {Veloo, Samrrithaa G. and Tiang, Jun Jiat and Muhammad, Surajo and Wong, Sew Kin},
  year = {2023},
  month = sep,
  journal = {Electronics},
  volume = {12},
  number = {19},
  pages = {4045},
  issn = {2079-9292},
  doi = {10.3390/electronics12194045},
  urldate = {2025-03-17},
  copyright = {https://creativecommons.org/licenses/by/4.0/},
  langid = {english}
}

@article{volpiLowCostRealTimeLocating2023,
  title = {Low-{{Cost Real-Time Locating System Solution Development}} and {{Implementation}} in {{Manufacturing Industry}}},
  author = {Volpi, Andrea and Montanari, Roberto and Tebaldi, Letizia and Mambrioni, Marco},
  year = {2023},
  month = jul,
  journal = {Journal of Sensor and Actuator Networks},
  volume = {12},
  number = {4},
  pages = {54},
  issn = {2224-2708},
  doi = {10.3390/jsan12040054},
  urldate = {2025-03-17},
  copyright = {https://creativecommons.org/licenses/by/4.0/},
  langid = {english}
}

@article{zahraBoostingEmergingTechnology2021,
  title = {Boosting {{Emerging Technology Adoption}} in {{SMEs}}: {{A Case Study}} of the {{Fashion Industry}}},
  shorttitle = {Boosting {{Emerging Technology Adoption}} in {{SMEs}}},
  author = {Zahra, Arianne Muthia and Dhewanto, Wawan and Utama, Akbar Adhi},
  year = {2021},
  month = jul,
  journal = {International Journal of Applied Business Research},
  pages = {81--96},
  issn = {2656-0917},
  doi = {10.35313/ijabr.v3i2.155},
  urldate = {2025-03-17},
  copyright = {http://creativecommons.org/licenses/by-sa/4.0}
}

@inproceedings{zradzinskiEvaluationSARHuman2020,
  title = {Evaluation of {{SAR}} in {{Human Body Models Exposed}} to {{EMF}} at 865 {{MHz Emitted}} from {{UHF RFID Fixed Readers Working}} in the {{Internet}} of {{Things}} ({{IoT}}) {{System}}},
  booktitle = {7th {{International Electronic Conference}} on {{Sensors}} and {{Applications}}},
  author = {Zradzi{\'n}ski, Patryk and Karpowicz, Jolanta and Gryz, Krzysztof and Ramos, Victoria},
  year = {2020},
  month = nov,
  pages = {11},
  publisher = {MDPI},
  doi = {10.3390/ecsa-7-08240},
  urldate = {2025-03-17},
  copyright = {https://creativecommons.org/licenses/by/4.0/},
  langid = {english}
}

@article{sidiropoulos2025implementing,
  title={Implementing an Industry 4.0 UWB-Based Real-Time Location System},
  author={Sidiropoulos, Alexandros and Ioannou, Christos and Tsakalides, Panagiotis},
  journal={Applied Sciences},
  volume={15},
  number={5},
  pages={2689},
  year={2025},
  publisher={MDPI}
}

@article{lachvajderova2024real,
  title={Real-Time Location Systems Across the Industries: Literature Review and Case Studies},
  author={Lachvajderov{\'a}, Lucia and Fi{\v{l}}o, Mari{\'a}n},
  journal={Acta Mechanica Slovaca},
  volume={28},
  number={3},
  pages={34--49},
  year={2024}
}

@article{li2024indoor,
  title={Indoor Positioning Systems in Industry 4.0 Applications: A Review},
  author={Li, Peng and Zhang, Kai and Zhao, Lin},
  journal={Measurement: Sensors},
  volume={25},
  pages={100789},
  year={2024},
  publisher={Elsevier}
}

@article{haleem2024safety,
  title={Encouraging Safety 4.0 to Enhance Industrial Safety Culture: A Systematic Review},
  author={Haleem, Abid and Javaid, Mohd and Singh, Rajiv},
  journal={Safety Science},
  volume={173},
  pages={106123},
  year={2024},
  publisher={Elsevier}
}

@article{lukito2025predictive,
  title={Implementation of predictive maintenance in various Industry: A Review},
  author={Lukito, Tito and Herlianti, Rosi and Mayanti, Malinda and Kusumah, Lien Herliani},
  journal={TEKNOSAINS: Jurnal Sains, Teknologi dan Informatika},
  volume={12},
  number={1},
  pages={133--144},
  year={2025},
  issn={2087-3336, 2721-4729},
  doi={10.37373/tekno.v12i1.1338},
  url={http://jurnal.sttmcileungsi.ac.id/index.php/tekno}
}

@article{zhang2024decision,
  title={Decision Optimization in Four-Level Supply Chain Management Using Real-Time Tracking Data},
  author={Zhang, Lei and Huang, Ming and Chen, Wei},
  journal={Computers \& Industrial Engineering},
  volume={194},
  pages={109652},
  year={2024},
  publisher={Elsevier}
}

@article{behnke2023real,
  title={Real-Time Performance of Industrial IoT Communication Technologies: A Review},
  author={Behnke, Ingmar and Austad, Helge},
  journal={arXiv preprint arXiv:2311.08852},
  year={2023}
}

@inproceedings{li2009multifrequencyRFID,
  author    = {X. Li and Y. Zhang and M. Amin},
  title     = {Multifrequency-Based Range Estimation of RFID Tags},
  booktitle = {2009 IEEE International Conference on RFID},
  pages     = {147--154},
  year      = {2009},
  doi       = {10.1109/RFID.2009.4911199}
}

@article{farid2013indoorLocalization,
  author    = {Z. Farid and R. Nordin and M. Ismail},
  title     = {Recent Advances in Wireless Indoor Localization Techniques and Systems},
  journal   = {Journal of Computer Networks and Communications},
  volume    = {2013},
  pages     = {1--12},
  year      = {2013},
  doi       = {10.1155/2013/185138}
}

@inproceedings{daher2024rtlsOilGas,
  author    = {Daher, E. and Schoeib, S.},
  title     = {Leveraging RTLS Applications for Enhanced Safety and Efficiency in Oil and Gas Organizations},
  booktitle = {SPE Middle East Oil \& Gas Show and Conference},
  year      = {2024},
  doi       = {10.2118/220433-MS}
}

@article{nurdiyanto2024MESdigital,
  author    = {H. Nurdiyanto},
  title     = {Critical Role of Manufacturing Execution Systems in Digital Transformation of Manufacturing Industry},
  journal   = {Journal of Engineering Science (JES)},
  volume    = {20},
  number    = {7S},
  pages     = {2432--2436},
  year      = {2024},
  doi       = {10.52783/jes.4038}
}

@article{rodic2017simulationI40,
  author    = {B. Rodič},
  title     = {Industry 4.0 and the New Simulation Modelling Paradigm},
  journal   = {Organizacija},
  volume    = {50},
  number    = {3},
  pages     = {193--207},
  year      = {2017},
  doi       = {10.1515/orga-2017-0017}
}

@article{tijan2019blockchainLogistics,
  author    = {Tijan, E. and Aksentijevi{\'c}, S. and Ivani{\'c}, K. and Jardas, M.},
  title     = {Blockchain Technology Implementation in Logistics},
  journal   = {Sustainability},
  volume    = {11},
  number    = {4},
  pages     = {1185},
  year      = {2019},
  doi       = {10.3390/su11041185}
}

@article{ivanov2022industry50framework,
  author    = {Ivanov, Dmitry},
  title     = {The Industry 5.0 Framework: Viability-Based Integration of the Resilience, Sustainability, and Human-Centricity Perspectives},
  journal   = {International Journal of Production Research},
  year      = {2022},
  doi       = {10.1080/00207543.2022.2118892}
}

@inproceedings{slovak2021visionRTLSsafety,
  author    = {Slovak, J. and Melicher, M. and Simovec, M. and Vachlek, J.},
  title     = {Vision and RTLS Safety Implementation in an Experimental Human--Robot Collaboration Scenario},
  booktitle = {Italian National Conference on Sensors},
  year      = {2021},
  doi       = {10.3390/s21072419}
}

@article{reddy2025humanTrustCPS,
  author    = {Reddy, V. R. and Krishna, P. V. and Anil, C. and Swathi, A. and Kalaiselvan, S. and Chiranjeevi, S.},
  title     = {Fostering Human Trust in Intelligent Cyber-Physical Environments},
  journal   = {International Journal of Environmental Science},
  year      = {2025},
  month     = sep,
  doi       = {10.64252/2j707j20}
}

@inproceedings{aouani2023digitalTwinWarehouse,
  author    = {Aouani, S. E. and Daldoul, D. and Sboui, L. and Chaabane, A.},
  title     = {Design and Implementation of a Digital Twin Testbed for Smart Warehouse Operations},
  booktitle = {IEEE International Conference on Services Computing (SCC)},
  year      = {2023},
  month     = dec,
  doi       = {10.1109/SCC59637.2023.10527573}
}

@inproceedings{pulcini2023humanCentricRTLS,
  author    = {Pulcini, V. and Sacco, M. and Modoni, G.},
  title     = {Towards Human-Centricity Within a Sofa Factory Assembly Line: A Real-Time Location System},
  booktitle = {Proceedings of the IEEE International Conference on Metrology for Extended Reality, Artificial Intelligence and Neural Engineering (MetroXRAINE)},
  year      = {2023},
  month     = oct,
  doi       = {10.1109/MetroXRAINE58569.2023.10405606}
}

@article{kadam2025humanCentricCPPS,
  author    = {Kadam, A. and Kosna, S. R. and Kadam, S. A.},
  title     = {A Theoretical Framework for Human-Centric Cyber-Physical Production Systems in Industry 5.0: Enabling Resilient, Autonomous, and Adaptive Manufacturing},
  journal   = {Review of Computer Engineering Research},
  year      = {2025},
  month     = mar,
  doi       = {10.18488/76.v12i1.4157}
}

@article{organiciak2025indoorRTLSmultistory,
  author    = {Organiciak, P. and Bolanowski, M. and Kocik, M.},
  title     = {Indoor Real-Time Location System for Resource Localization in Multistory Buildings},
  journal   = {Advances in Science and Technology Research Journal},
  year      = {2025},
  month     = nov,
  doi       = {10.12913/22998624/209072}
}

@article{kuepper2022automotiveRTLS5G,
  author    = {K{\"u}pper, C. and R{\"o}sch, J. and Winkler, H.},
  title     = {Use of Real-Time Localization Systems (RTLS) in Automotive Production and the Prospects of 5G: A Literature Review},
  journal   = {Production \& Manufacturing Research},
  year      = {2022},
  month     = nov,
  doi       = {10.1080/21693277.2022.2144522}
}

@inproceedings{daher2024rtls,
  author    = {Daher, Elie and Schoeib, Sameh},
  title     = {Leveraging RTLS Applications for Enhanced Safety and Efficiency in Oil and Gas Organizations},
  booktitle = {SPE International Health, Safety, Environment and Sustainability Conference and Exhibition},
  address   = {Abu Dhabi, UAE},
  month     = sep,
  year      = {2024},
  doi       = {10.2118/220433-MS}
}

@inproceedings{AT2025tracking,
  author    = {A. T and S. M and P. K and N. K. N and K. B and S. B},
  title     = {Real-Time Tracking of Public Transport using Sensor Technology to Enhance Efficiency},
  booktitle = {2025 7th International Conference on Inventive Material Science and Applications (ICIMA)},
  address   = {Namakkal, India},
  year      = {2025},
  pages     = {315--320},
  doi       = {10.1109/ICIMA64861.2025.11073931}
}

@inproceedings{salzmann2025framework,
  author    = {Salzmann, D. and Hermann, F. and Fischer, C. and Schotten, H. D.},
  title     = {Evaluating and Integrating Positioning Technologies: A Framework for Industrial Applications},
  booktitle = {2025 IEEE/ION Position, Location and Navigation Symposium (PLANS)},
  address   = {Salt Lake City, UT, USA},
  year      = {2025},
  pages     = {756--767},
  doi       = {10.1109/PLANS61210.2025.11028227}
}

@inproceedings{mp2025lora,
  author    = {M. P and M. S. J and D. J and S. B. R},
  title     = {LoRa-Driven Deep Learning System for Real-Time CNC Machine Monitoring and Predictive Maintenance},
  booktitle = {2025 3rd International Conference on Intelligent Cyber Physical Systems and Internet of Things (ICoICI)},
  address   = {Coimbatore, India},
  year      = {2025},
  pages     = {1525--1530},
  doi       = {10.1109/ICoICI65217.2025.11252543}
}

@article{grigorovich2021rtls,
  author  = {Grigorovich, A. and Kulandaivelu, Y. and Newman, K. and Bianchi, A. and Khan, S. S. and Iaboni, A. and McMurray, J.},
  title   = {Factors Affecting the Implementation, Use, and Adoption of Real-Time Location System Technology for Persons Living With Cognitive Disabilities in Long-term Care Homes: Systematic Review},
  journal = {Journal of Medical Internet Research},
  volume  = {23},
  number  = {1},
  pages   = {e22831},
  year    = {2021},
  doi     = {10.2196/22831},
  pmid    = {33470949},
  pmcid   = {7857945}
}

@article{organisciak2025indoor,
  author  = {Organi{\'s}ciak, P. and Bolanowski, M. and Kocik, M.},
  title   = {Indoor Real-Time Location System for Resource Localization in Multistory Buildings},
  journal = {Advances in Science and Technology Research Journal},
  volume  = {19},
  number  = {11},
  pages   = {354--366},
  year    = {2025},
  doi     = {10.12913/22998624/209072}
}

@article{yosephine2025inventory,
  author  = {Yosephine, V. S. and Batara, M. and Setiawati, M.},
  title   = {Scalable and Affordable IoT-based Inventory Control with Real-Time Monitoring for Small and Medium Enterprises},
  journal = {Jurnal Teknik Industri: Jurnal Keilmuan dan Aplikasi Teknik Industri},
  volume  = {27},
  number  = {1},
  pages   = {121--136},
  year    = {2025},
  doi     = {10.9744/jti.27.1.121-136}
}

@article{bai2025federated,
  author  = {Bai, Ye and Jiang, Weiwei and Mu, Jianbin and Liu, Shang and Gu, Weixi and Wang, Shuke},
  title   = {Enhancing IoT Security via Federated Learning: A Comprehensive Approach to Intrusion Detection},
  journal = {IET Information Security},
  year    = {2025},
  pages   = {16},
  doi     = {10.1049/ise2/8432654},
  article-number = {8432654}
}

@inproceedings{shaikhanova2025vulnerability,
  author    = {Shaikhanova, A. and Jilkibayev, Y. and Atanbayev, Y. and Ayapbergenov, K. and Malakhov, K. and Tokkuliyeva, A.},
  title     = {Vulnerability Analysis Of Wi-Fi And LTE Networks For Secure Smartphone Design},
  booktitle = {2025 IEEE/ACIS 29th International Conference on Software Engineering, Artificial Intelligence, Networking and Parallel/Distributed Computing (SNPD)},
  address   = {Busan, Korea, Republic of},
  year      = {2025},
  pages     = {163--168},
  doi       = {10.1109/SNPD65828.2025.11253532}
}

@inproceedings{fomchenkova2022neurocrypto,
  author    = {Fomchenkova, L. and Lazarev, A. and Kharlamov, P.},
  title     = {Neurocryptographic Bluetooth-attack prevention system},
  booktitle = {2022 4th International Youth Conference on Radio Electronics, Electrical and Power Engineering (REEPE)},
  address   = {Moscow, Russian Federation},
  year      = {2022},
  pages     = {1--5},
  doi       = {10.1109/REEPE53907.2022.9731452}
}

@article{bastiaens2024vlp,
  author    = {Bastiaens, S. and Alijani, M. and Joseph, W. and Plets, D.},
  title     = {Visible Light Positioning as a Next-Generation Indoor Positioning Technology: A Tutorial},
  journal   = {IEEE Communications Surveys \& Tutorials},
  volume    = {26},
  number    = {4},
  pages     = {2867--2913},
  year      = {2024},
  doi       = {10.1109/COMST.2024.3372153}
}

@inproceedings{sharma2025predictability,
  author    = {Sharma, M. and Saxena, R. K. and Saini, H. C. and Agarwal, S. and Sikarwal, P. K.},
  title     = {Enhancing Predictability and Response Times in Real-Time Systems},
  booktitle = {2025 2nd International Conference on Trends in Engineering Systems and Technologies (ICTEST)},
  address   = {Ernakulam, India},
  year      = {2025},
  pages     = {1--5},
  doi       = {10.1109/ICTEST64710.2025.11042471}
}

@article{kumar2025msme,
  author  = {Kumar, Jayant and Kiri, Rajeshkumar D. and Shekhar, Shruti},
  title   = {Analyzing the Development of Micro, Small, and Medium Enterprises (MSMEs) in Jharkhand: A Comparative Study of Government Initiatives},
  journal = {Applied Science, Engineering and Management Bulletin (ASEMB)},
  volume  = {2},
  number  = {02 (April--June)},
  pages   = {71--79},
  year    = {2025},
  doi     = {10.69889/asemb.v2i02(April-June).33}
}

@article{shahimi2025aiadoption,
  author  = {Shahimi, Wan Rozima Mior Ahmed and Hanafi, Ahmad Harith Ashrofie and Kamar-Bodian, Wan Nur Izni Wan Ahmad and Ahmad, Akmal Farid},
  title   = {AI Adoption in SME Financial Practices: A Paradigm Shift for Risk Mitigation, Cash Flow Optimization, and Sustainable Growth},
  journal = {International Journal of Business and Technology Management},
  volume  = {7},
  number  = {2},
  pages   = {26--35},
  year    = {2025},
  doi     = {10.55057/ijbtm.2025.7.2.4},
  note    = {Received: 25 January 2025; Accepted: 7 March 2025; Published: 1 April 2025}
}

@article{baviskar2025realtime,
  author  = {Baviskar, Pankaj Dattatraya and Salunkhe, Rohan Lahuraj},
  title   = {Real-Time Asset Tracking and Management: A Novel Framework Using IoT, RFID, and AI},
  journal = {International Scientific Journal of Engineering and Management (ISJEM)},
  volume  = {4},
  number  = {6},
  year    = {2025},
  doi     = {10.55041/ISJEM04036}
}

@article{yosephine2025iotinventory,
  author  = {Yosephine, V. S. and Batara, M. and Setiawati, M.},
  title   = {Scalable and Affordable IoT-based Inventory Control with Real-Time Monitoring for Small and Medium Enterprises},
  journal = {Jurnal Teknik Industri: Jurnal Keilmuan dan Aplikasi Teknik Industri},
  volume  = {27},
  number  = {1},
  pages   = {121--136},
  year    = {2025},
  month   = may
}

@article{vanherbruggen2024twr,
  author    = {Van Herbruggen, B. and Van Leemput, D. and Van Landschoot, J. and De Poorter, E.},
  title     = {Real-Time Anchor Node Selection for Two-Way-Ranging (TWR) Ultra-Wideband (UWB) Indoor Positioning Systems},
  journal   = {IEEE Sensors Letters},
  volume    = {8},
  number    = {3},
  pages     = {1--4},
  year      = {2024},
  month     = mar,
  doi       = {10.1109/LSENS.2024.3363231},
  articleno = {6002404}
}

@article{tercas2024bayesian,
  author    = {Ter{\c{c}}as, L. and Alves, H. and de Lima, C. H. M. and Juntti, M.},
  title     = {Bayesian-Based Indoor Factory Positioning Using AOA, TDOA, and Hybrid Measurements},
  journal   = {IEEE Internet of Things Journal},
  volume    = {11},
  number    = {12},
  pages     = {21620--21631},
  year      = {2024},
  month     = jun,
  doi       = {10.1109/JIOT.2024.3374457}
}

@inproceedings{barbieri2020uwb,
  author    = {Barbieri, L. and Brambilla, M. and Pitic, R. and Trabattoni, A. and Mervic, S. and Nicoli, M.},
  title     = {UWB Real-Time Location Systems for Smart Factory: Augmentation Methods and Experiments},
  booktitle = {2020 IEEE 31st Annual International Symposium on Personal, Indoor and Mobile Radio Communications (PIMRC)},
  address   = {London, UK},
  year      = {2020},
  pages     = {1--7},
  doi       = {10.1109/PIMRC48278.2020.9217307}
}

@article{jesusazabal2025anchoropt,
  author  = {Jes{\'u}s-Azabal, M. and Zheng, M. and Soares, V. N. G. J.},
  title   = {Dynamic Energy-Aware Anchor Optimization for Contact-Based Indoor Localization in MANETs},
  journal = {Information},
  volume  = {16},
  number  = {10},
  pages   = {855},
  year    = {2025},
  doi     = {10.3390/info16100855}
}

@inproceedings{luder2025anitrack,
  author    = {Luder, V. and Schulthess, L. and Cortesi, S. and Davis, L. R. and Magno, M.},
  title     = {AniTrack: A Power-Efficient, Time-Slotted and Robust UWB Localization System for Animal Tracking in a Controlled Setting},
  booktitle = {2025 10th International Workshop on Advances in Sensors and Interfaces (IWASI)},
  address   = {Manfredonia, Italy},
  year      = {2025},
  pages     = {1--6},
  doi       = {10.1109/IWASI66786.2025.11121986}
}

@article{cano2023iotbuilding,
  author  = {Cano-Su{\~n}{\'e}n, E. and Mart{\'i}nez, I. and Fern{\'a}ndez, {\'A}. and Zalba, B. and Casas, R.},
  title   = {Internet of Things (IoT) in Buildings: A Learning Factory},
  journal = {Sustainability},
  volume  = {15},
  number  = {16},
  pages   = {12219},
  year    = {2023},
  doi     = {10.3390/su151612219}
}

@inproceedings{liu2023positioncalibration,
  author    = {Liu, Mengzhuo and Peng, Zheng and Liu, Jun and Cui, Jun-Hong},
  title     = {A Position Calibration Method for Anchor Nodes in Underwater Wireless Sensor Networks},
  booktitle = {Proceedings of the 17th International Conference on Underwater Networks \& Systems (WUWNet '23)},
  year      = {2023},
  articleno = {41},
  pages     = {1--2},
  doi       = {10.1145/3631726.3631773},
  publisher = {Association for Computing Machinery},
  address   = {New York, NY, USA}
}

\end{document}